\documentstyle[amssymb,aps,prl,psfig,twocolumn]{revtex}
\begin{document}
\draft


\title{Rydberg molecules in external fields: a semiclassical analysis}

\author{A.\ Matzkin and T. S. Monteiro}

\address{Department of Physics and Astronomy, University College London,
Gower Street, London WC1E 6BT, U.K.}

\date{\today}

\maketitle

\begin{abstract}
We undertake a semiclassical analysis of the spectral properties
(modulations of photoabsorption spectra, energy level statistics) of a
simple Rydberg molecule in static fields within the framework of
Closed-Orbit/Periodic-Orbit theories. We conclude that in addition to the
usual classically allowed orbits one
must consider classically forbidden diffractive paths. Further, the molecule
brings in a new type of 'inelastic' diffractive trajectory, different from
the usual 'elastic' diffractive orbits encountered in previous studies of
atomic and analogous systems such as billiards with point-scatterers. The
relative importance of inelastic versus elastic diffraction is quantified by
merging the usual Closed Orbit theory framework with molecular quantum
defect theory.

\end{abstract}

\pacs{32.60.+i, 05.45.-a, 03.65.Sq}

Gutzwiller Periodic Orbit theory (POT) \cite{gutzwiller90}
represents a very successful technique for
establishing a correspondence between quantum properties and classically
chaotic dynamics. Recently there has been much interest in {\em diffractive}
POT, e.g. \cite{Vattay}, for systems with a
dynamical structure smaller or comparable to one de Broglie wavelength.
These studies were restricted to a structureless point-scatterer,
such as a simple $s$-wave scatterer. It is an
interesting question, from the point of view of quantum chaology, to consider
the effect of a diffractive scatterer with internal structure, such as for
example a two-level particle. In fact we show here that this model provides
us with the key to analyze semiclassically, for the first time, the spectra
of diatomic Rydberg molecules such as H$_{2}$ in static fields. We find that
features such as the interference between Rydberg series converging on
different energy thresholds depend on a novel type of {\em inelastic}
diffractive trajectory.

 Highly excited hydrogen atoms in static magnetic
fields represents a well-known paradigm of quantum chaology: modulations in
the quantum density of states are well described by POT \cite{Wintg90}.
 However, analysis of the {\em photoabsorption}  spectra from a
low-lying initial state, require Closed-orbit theory (COT) \cite{delos88}, a semiclassical description of a density of states
weighted by a dipole matrix element. Originally devised for
hydrogen, COT was later extended to singly excited
non-hydrogenic ('Rydberg') atoms ~\cite{delos88,dando etal95},
accounting for the presence of the ionic core. It was found ~\cite
{dando etal95} that the core-scattering 
produces additional 'combination' orbits. Non-hydrogenic
energy-level spectra, even in weak fields where the classical dynamics
for hydrogen is near-integrable, were found to be significantly more complex, showing short-range level repulsion and a proliferation of spectral
modulations, a behavior termed 'core-induced chaos'. Subsequently,
 the nearest neighbor spacings (NNS)
distributions of non-hydrogenic atoms in the 'core-induced chaos' regime
were shown in fact \cite{jonckheere etal98}, to correspond to a distribution
which was neither Wigner (chaotic), nor mixed phase-space, but close to a
new generic {\it intermediate} class known as {\em semi-Poisson} \cite
{bogomolny etal99}. These findings were investigated experimentally for
helium atoms in external fields \cite{karremans etal99}. 

We consider below the form of COT appropriate for
 simple molecules in external fields. The extension of COT is obtained by merging the standard closed-orbit
arguments with molecular quantum defect theory (MQDT), the main theoretical
approach employed in the studies of Rydberg molecules \cite{jungen ed96}.
The success of MQDT is based on two ingredients: the separation of the
core-electron interaction into long-range effects (described by Coulomb
functions) and short-range effects (described by phase-shifts known as {\em %
quantum defects}), and the use of different basis sets to account for
core-electron couplings. Thus, when the electron is near the core, its
dynamics is coupled to the rotation of the core, the wavefunction being
described in the coupled frame $\left| \alpha \right\rangle \equiv \left|
\Lambda _{\alpha }l_{\alpha }J_{\alpha }\right\rangle $, ($\Lambda $ is the
projection of the electronic angular momentum on the molecular axis, $l$ is
the orbital momentum of the outer electron, quantized along this axis, and $%
J $ gives the total angular momentum). For each $\alpha ,$ the quantum
defect is given by the number $\mu _{\alpha }$ and scattering matrices have
a simple representation, e.g. the $T$ matrix elements are $T_{\alpha
}=e^{2i\pi \mu _{\alpha }}-1$. When the electron is far from the core, the
molecular wavefunction is described in the uncoupled frame $\left|
j\right\rangle \equiv \left| N_{j}l_{j}m_{j}\right\rangle $ ($N$ is the
angular momentum of the freely rotating core, $l$ is now quantized in the
laboratory frame along the axis of the magnetic field with magnetic quantum
number $m)$. These two coupling schemes are related by a unitary frame
transformation with elements $\left\langle j\right| \left. \alpha
\right\rangle $. The $T$ matrix elements between outer channels $j$ and $%
j^{\prime }$are given in terms of $T_{\alpha }$ by $T_{jj^{\prime
}}=\sum_{\alpha }\left\langle j\right. \left| \alpha \right\rangle T_{\alpha
}\left\langle \alpha \right. \left| j^{\prime }\right\rangle .$ The frame
transformation in the {\em field-free} case was given by the late
U.\ Fano in a classic paper \cite{fano70} which successfully analyzed
experimental photoabsorption spectra for H$_{2}$ from the ground state ($%
J=0, $ $l=0$) to high Rydberg states ($J=1,$ $l=1$). In the presence of a
magnetic field of strength $\gamma $, the frame transformation is modified
since $J$ is not conserved, but $M,$ its projection on the field axis, is 
\cite{monteiro taylor90}.

The quantum MQDT treatment, undertaken in an inner region where the magnetic
field is negligible, is matched to the semiclassical one in the following way
(see Fig. 1). (1) The molecule, initially in its ground state,
 is excited by a laser, giving waves with $l=1,$ $%
J=1,$ and both $\Lambda =0$ (phase-shifted by the quantum defect $\mu
_{\Sigma }$) and $\Lambda =1$ (with a phase-shift given by $\mu _{\Pi }$).
(2) As the electron leaves the core region, its dynamics gets uncoupled from
the core which is left in one of the 4 allowed configurations: $N=0$ $m=0$
and $N=2$ $m=0,\pm 1$. These excited waves are propagated by the full
field-free quantum molecular Green's function, obtained from the
Lippmann-Schwinger form of the MQDT wavefunction \cite{matzkin99}.

 (3) The
electron then enters the region in which the magnetic field is not
negligible, where the total wavefunction takes the form 
$ \left| \psi \right \rangle =\sum_{j} \sum_{N_{j}}\left|
N_{j}M-m_{j}\right \rangle e^{im_{j}\phi }\psi _{out}^{m_{j}N_{j}} $.
 Here, $\psi _{out}^{m_{j}N_{j}}$ is similar to the outgoing wave of the electron
in COT \cite{delos88}. For {\em each} core configuration, the electron's
wavefunction is propagated in the magnetic field region semiclassically
along classical trajectories $k$. Two dynamical regimes coexist, since the
energy partition gives 
\begin{equation}
E_{N=0}^{el}=E_{N=2,m}^{el}+6B_{r}+m\gamma /2  \label{2}
\end{equation}
where $E^{el}$ is the outer electron's energy and $B_{r}$ the core
rotational constant (the dependence on $m$ through the Zeeman term for the $%
N=2$ levels is very small). (4) Eventually some trajectories return to the
core. The molecular wavefunction is then written as a standard MQDT expansion,
and the expansion coefficients may be obtained 
 by matching the incoming part of the wavefunction to the
semiclassical returning wave. (5) The
electron is then projected onto the coupled frame, and the returning waves
interfere with the initially excited waves to produce modulations in the
absorption spectra.

These modulations contain contributions from the
primitive closed orbits $k$ that left and returned to the core in one of its
4 possible configurations $j$ and are obtained by summing terms of the form
\begin{equation} 
 \lbrack \sum_{\alpha ,\alpha ^{\prime }}{\cal C}_{\alpha }{\cal C}_{\alpha
^{\prime }}\left\langle \alpha \right| \left. j\right\rangle \left\langle
j\right| \left. \alpha ^{\prime }\right\rangle e^{i\pi (\mu _{\alpha }+\mu
_{\alpha ^{\prime }})}]{\cal A}_{k}^{j}e^{iS_{k}^{j}} \label{4}
\end{equation}
where ${\cal A}_{k}^{j}$ and $S_{k}^{j}$ are the classical amplitude and
action for trajectory $k$ and energy $E_{j}$ ($S_{k}$ includes the associated
semiclassical phase, containing the Maslov index); ${\cal C}_{\alpha }=1$
if $\Lambda =0$ and $\sqrt{2}$ if $\Lambda =1$ is a coefficient stemming
from the use of the united atom approximation for the dipole transitions.
(6) The Coulomb field backscatters the returning waves which retrace the orbit in reverse, giving
rise to repetitions without changing the state of the core. (7) Core-scattering produces newly outgoing waves, which are propagated in the outer region
along orbits $q$. These waves now leave the core in state $N_{j}m_{j}$ but
contain contributions from orbits that had previously returned with the core
being in state $N_{j^{\prime }}m_{j^{\prime }}$, with the mixing strengths
depending both on the classical dynamics (via the expansion coefficients) and the quantum scattering process (via the $T$ matrix elements). These
orbits return to the core for the second time where they are matched once
again to a MQDT expansion. The contribution of these orbits to the
oscillatory part of the photoabsorption spectrum is of the form 
\begin{eqnarray}
&&[\sum_{\alpha ,\alpha ^{\prime }}{\cal C}_{\alpha }{\cal C}_{\alpha
^{\prime }}\left\langle \alpha \right| \left. j\right\rangle \left\langle
j\right| \left. \alpha ^{\prime }\right\rangle e^{i\pi (\mu _{\alpha }+\mu
_{\alpha ^{\prime }})}T_{jj^{\prime }}]  \nonumber \\
&&\times {\cal A}_{j}^{q}{\cal A}_{j^{\prime }}^{k}e^{i\left(
S_{q}^{j}+S_{k}^{j^{\prime }}\right) }  \label{5}
\end{eqnarray}
which clearly singles out the combination of an orbit $k$ at an energy $%
E_{j^{\prime }}$ with an orbit $q$ having an energy $E_{j}$. Further
iterations may be performed, but since each process of matching the incoming
wave to the MQDT expansion brings in a factor $\hbar ^{1/2},$ fast
convergence is expected for small $\hbar $.

COT for atoms has relied on exact or quasi-exact scaling properties since the
classical dynamics of the excited electron depends only on a scaled
energy $\epsilon =E\gamma ^{-2/3}$. The field represents an effective
value of Planck's constant $\hbar = \gamma ^{1/3}$. Hydrogen in a magnetic field is
near-integrable for $\epsilon <-0.5$; as the field is increased it makes a
gradual transition to full chaos at $\epsilon \simeq -0.1$. Although here the
dynamics associated with the core in states $N=0$ and 2 cannot be scaled
simultaneously, scaling techniques may still be
applied theoretically to unravel the classical dynamics
 (e.g. \cite{main etal97}).
Eq. (\ref{2}) is effectively scaled by giving a $\gamma ^{2/3}$ dependence
to the rotational constant thereby allowing calculations at fixed scaled
energies $\epsilon _{N=0}$ {\em and} $\epsilon _{N=2}$. Full quantal
calculations of the photoabsorption spectra were obtained by scaling the
method described in \cite{monteiro taylor90,he etal96}. Note that $l$-mixing
brings in (very weakly) the effects of $N=4,6...$ states. However their
contribution is very small and although we include these in our quantal
calculations, the physics remains in effect dominated by the $N=0,2$ series
of quantum states. The H$_2$ molecule and its isotopes D$_2$ and T$_2$
have similar quantum defects. We present the results for T$_2$ below since
its $B_{r}$ yields the smallest values of $\hbar$ in both channels 
($\hbar \sim 1/90$ and $\sim 1/40$ for $N=0$ and $N=2$
respectively).

The Fourier transforms of the spectra are shown in Fig. 2 for $M=0$ at a
scaled energy $\epsilon _{N=0}=-0.3,$ 
corresponding to $\epsilon_{N=2}=-2.45,$ over a 
range $\gamma ^{-1/3}=60-120.$ Over this range, $B_{r}$
varies from about twice to half its physical value. Each peak is associated
with a closed orbit, either a 'geometric' orbit $k$ (i.e. one of 
the well-known \cite{holle etal88} closed orbits of the diamagnetic hydrogen
atom problem)  appearing
at scaled action $\widetilde{S}_{k}$ or a 
'diffractive' combination of at least two orbits $k$ and $%
q $  at actions $\widetilde{S}_{k}+%
\widetilde{S}_{q}$. 
When $\mu _{\Sigma }=\mu _{\Pi }$, the 
terms between brackets in Eqs. (\ref
{4}) and (\ref{5}) vanish for $N_{j}=2$ and $m_{j}=0,\pm 1$ (for any $%
j^{\prime }$) hence only peaks associated with the  ($%
\epsilon _{N=0}$) closed orbits are expected. For zero quantum
defects [Fig. 2(a)] there is no core scattering (the $T$ matrix vanishes)
and only geometric orbits are visible. For 
$\mu _{\Sigma }=\mu _{\Pi }=0.5$ [Fig. 2(b)], {\em %
elastic} core scattering between $\epsilon _{N=0}$ orbits produces the
additional peaks; these combination orbits were seen in
non-hydrogenic atomic spectra \cite{dando etal95}. (c) shows $\mu
_{\Sigma }=0.5,$ $\mu _{\Pi }=-0.5$, in which case Eqs. (\ref{4}) and (\ref
{5}) predict, as a generic feature, the geometric and elastic-scattered peaks
corresponding to $\epsilon _{N=0}$ to be about 10 times
weaker than the ones associated with the  $\epsilon _{N=2}$
closed orbits.\ Moreover, the term given in Eq. (\ref{5}) vanishes if $%
N_{j}=0$ and $N_{j^{\prime }}=2,m_{j^{\prime }}=0,\pm 1,$ totally
suppressing {\em inelastic} scattering. Indeed, the regularly spaced peaks
in Fig. 2(c), are associated with the closed orbits of the near integrable
$\epsilon_{N=2}$  case (the orbits perpendicular and parallel to the field for
$\epsilon_{N=2}=-2.45$ have quasi-degenerate actions) and their
repetitions. Thus Figs. 2(a)-2(c) display simple features characterizing
systems in which at most only elastic core-scattering is visible in the
photoabsorption spectrum.

In contrast, the Fourier spectra shown in Figs. 2(d)-(e) contain far more
peaks. These arise from {\em inelastic} scattering between the
core and the outer electron and appear at a scaled action corresponding to
the sum of an $\epsilon _{N=2}$ orbit (or its repetitions) and an $\epsilon
_{N=0}$ orbit.  The shortest of these combination orbits
occurs at $\widetilde{S}=1.71$ and combines the $V_{1}^{1}$ 
'balloon orbit' ~\cite
{dando etal95} at $\epsilon =-0.3$ [peak '1' in Fig. 2(a)] with
the  primitive orbit 'i' at $\epsilon =-2.45$ [the first peak in Fig. 2(c)].
Note that two or more peaks having the same action but arising from
different combinations will interfere constructively or destructively depending on the
relative phases which contain both quantum scattering and classical orbit
terms. However generic, peak-independent features are governed solely
by the quantum defects. For example, when 
$\mu _{\Sigma }=0.5,$ $\mu _{\Pi }=0,$
the term between brackets in Eq. (\ref{5}) is almost negligible when $%
N_{j}=N_{j^{\prime }}=0$, thereby effectively suppressing elastic
scattering.\ This explains why in Fig. 2(d) there are no peaks corresponding
to elastic core-scattering of the $\epsilon _{N=0}$ 
orbits. Fixing $\mu _{\Sigma }=0.5,$ the height of these elastic peaks 
 decreases monotonically when $\mu _{\Pi }$ is decreased from $%
0.5$ to $0$, as expected by our semiclassical formulas. For $\mu _{\Sigma
}=0.22,$ $\mu _{\Pi }=-0.06$ [Fig. 2(e)], the quantum amplitudes linking the
different channels have comparable magnitudes, and both elastic as well as
inelastic combinations are seen.

We stress that the classical dynamics underlying Figs. 2(a)-(e) 
 is identical, and the very different plots  confirm the importance
 of the type of core scattering in the
 recurrence spectra. For example the CO molecule has equal quantum defects and its
recurrence spectra should be similar to Fig. 2(b)
(elastic scattering only). Fig. 2(e) on the other
hand corresponds to H$_{2}$ which has the well-known quantum defects $\mu
_{\Sigma }=0.22,$ $\mu _{\Pi }=-.06$ at equilibrium internuclear distance.

We have also calculated the NNS distributions for different values of the
quantum defects at $\epsilon _{N=0}=-0.5$. The results are given on Fig. 3.
The distribution for $\mu _{\Sigma }=\mu _{\Pi }=0$ [Fig. 3(a)] appears as a
superposition of 2 Poisson distributions, the first one concentrated at
short-spacings ($s\simeq 0.3$), the second one extending its tail at longer
range. Note that although the Rydberg
series do not interact, the overall spacing distribution {\em is not}
equivalent to a superposition of 2 independent spectra (in which case the resulting distribution is well-known \cite{guhr etal98}). Nevertheless, it was possible to obtain an acceptable fit to the
phenomenological formula $
P(s)=\frac{\rho _{1}}{\rho }\left( \frac{\lambda }{\lambda -\rho _{2}}%
\right) P_{1}(\lambda ,s)+\frac{\rho _{2}}{\rho }\left( \frac{\lambda -\rho 
}{\lambda -\rho _{2}}\right) P_{2}(\rho _{2},s)$, with $\sum_{i}\rho _{i}=\rho $ where $\rho $ is the total
density. $P_{1}$ gives the dominant distribution at short-spacings and vanishes for $%
s\gtrsim 1,$ whereas $P_{2}$ gives the distribution which extends at larger
spacings, with mean spacing $\rho _{2}^{-1}$. $\lambda $ is treated as a fitting
parameter with $\lambda \geq \rho $. Thus for small spacings, $P(s)$ describes the
superposition of $P_{1}$ and $P_{2}$ with the limiting cases $P(s)\rightarrow P_{1}(\rho ,s)$ when 
$\lambda \rightarrow \rho$, and $P(s)\simeq \delta (s-\lambda ^{-1})\rho_{1}/\rho +P_{2}(\rho _{2},s)\rho _{2}/\rho$ when $\lambda \rightarrow \infty.$

$P_{1}(s)$ was always taken to be a Poisson
distribution, with $\lambda \simeq 4$. This is the consequence of the weak
field experienced by the $N=2$ levels, which remain clustered around their
field-free position. However, $P_{2}(s)$ does depend on the scatterer properties,
i.e. on the quantum defects. For $\mu _{\Sigma }=\mu _{\Pi }=0,$ $P_{2}$
follows a Poisson statistics (the tail of the solid line given in Fig.
3(a)). For $\mu _{\Sigma }=\mu _{\Pi }=0.5$ (b) the tail is in
excellent agreement with the Wigner surmise, as is also the case for $\mu
_{\Sigma }=0.5,$ $\mu _{\Pi }=-0.5$ (c). On the other hand, for $\mu
_{\Sigma }=0.5,$ $\mu _{\Pi }=0$ [Fig. 3(d)] and for $\mu _{\Sigma }=0.22,$ $%
\mu _{\Pi }=-0.06$ (e), $P_{2}$ is best approximated by a
semi-Poisson distribution. Globally, $P_{2}$ goes from Poisson to Wigner,
through intermediate statistics, as the intensity of the scatterer is
increased. This intensity is given strictly by a multidimensional matrix
 obtained from the frame transformation as
explained above [roughly, the intensity goes from $0$ to a maximum value
as $(\left| \mu _{\Sigma }\right| ,\left| \mu _{\Pi }\right| )$ goes from $%
(0,0)$ to $(0.5,0.5)$]. With our simple fit, the 
specific effects of inelastic scattering are not quantified. As
known, 2\ Rydberg series strongly interact when the levels cross each
other, and we may thus expect the inelastic effects to appear at rather
short spacings. We do
remark however a smaller value of $\lambda $ for the cases (d) and (e), where inelastic scattering is stronger, which indicates a
larger mean level repulsion for the distribution modelled by $P_{1}$.

In conclusion, we have shown that the relative contribution
of geometric and diffractive trajectories produced by
{ \em elastic and/or  inelastic} scattering on the core are essential 
to understand the character of recurrence
spectra of different simple molecules in external fields. The spectral properties for these molecular systems,
containing an effective 2-level scatterer, are dominated by 2 spacing
distributions.

\begin{figure}[ht]
\caption {Simplified view of the dynamics of photoabsorption of H$_2$ in a magnetic field. 
The numbered steps are detailed in the text, but in brief: (1) A ground-state 
diatomic molecule absorbs a photon. 
(2) Outgoing waves 
with $l=1$ leave the core region; the dynamics of the corresponding
excited electron uncouple from the core 
leaving it in a superposition of rotationally excited ($N=2$) or not 
excited ($N=0$) states. 
(3) In the extended region outside the core, the outgoing electron dynamics 
is described using 
classical trajectories. (4) Some trajectories return to the core 
carrying returning waves. (5) The returning waves interfere with the 
outgoing waves producing modulations in the spectra. Returning waves 
are either (6) back-scattered by the Coulomb field ('geometric' scattering
in the terminology of diffractive POT) 
producing higher repetitions (7), or are core-scattered 
('diffractive' scattering) without exchange of energy with the core 
(elastically) or with exchange (inelastically). The latter produce 
spectral modulations due to combinations of trajectories 
with different scaled energies (e.g. $\epsilon _{N=0}$ and $\epsilon _{N=2}$)
 and different classical regimes. They are not seen in 
 atomic spectra and are a new feature, seen only in the molecular spectra.
 }
\label{Fig.1}
\end{figure}

\begin{figure}[tb]
\caption {Fourier transformed quantum spectra  
for a set of diatomic molecules characterized 
by different quantum defects
$\mu _{\Sigma },\mu _{\Pi }$, showing the relative importance
of elastic and inelastic diffractive trajectories. 
(a) $\mu _{\Sigma },\mu _{\Pi } =0,0$:
no core scattering; peaks correspond to $N=0$ series (molecular  
core not
rotationally excited / outer electron highly ($n \sim 90$) excited)
 hence the COs
correspond to those of a hydrogen $atom $ at $\epsilon=-0.3$.
(b) Shows only {\em elastic} core-scattering, within the $N=0$ series.
(c) Spectra dominated by COs of $N=2$ states (molecular core rotationally
excited / low excitation ($n \sim 40$) electron) and elastic core
 scattering combining these trajectories.
(d) Shows trajectories of both $N=0$ and $N=2$ states. New core scattered
peaks are strongly dominated by {\em inelastic}
 diffractive trajectories combining 
$N=0$ and $N=2$ trajectories, i.e. the outer electron exchanges
 energy with the
 core. (e) $\mu _{\Sigma },\mu _{\Pi }$ correspond to H$_2$; shows mixture 
 of elastic and inelastic
 core-scattered orbits.}
\label{Fig.2}
\end{figure}

\begin{figure}[tpb]
\caption {Spacing distributions (NNS) for different 
values of the quantum defects [linear scale for (a), log scale 
for (b)-(e)]. The solid line represents the fit $P(s)$.
}\label{Fig.3}
\end{figure}

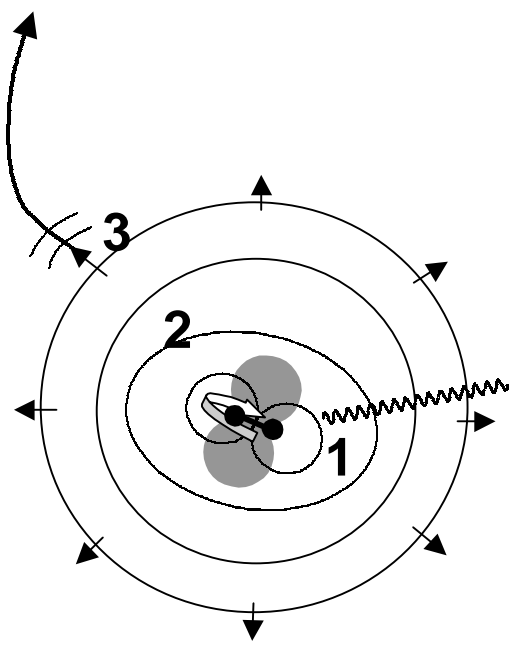
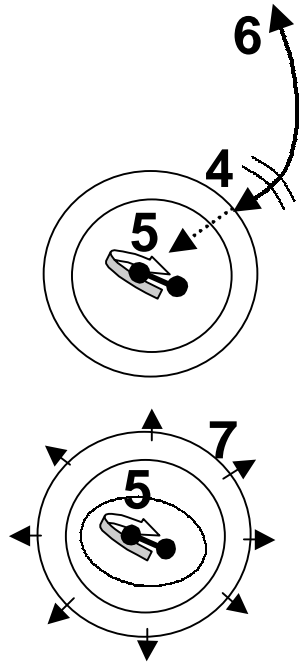
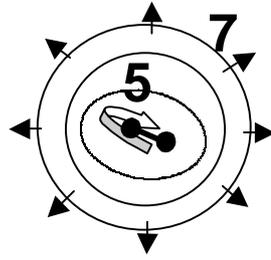
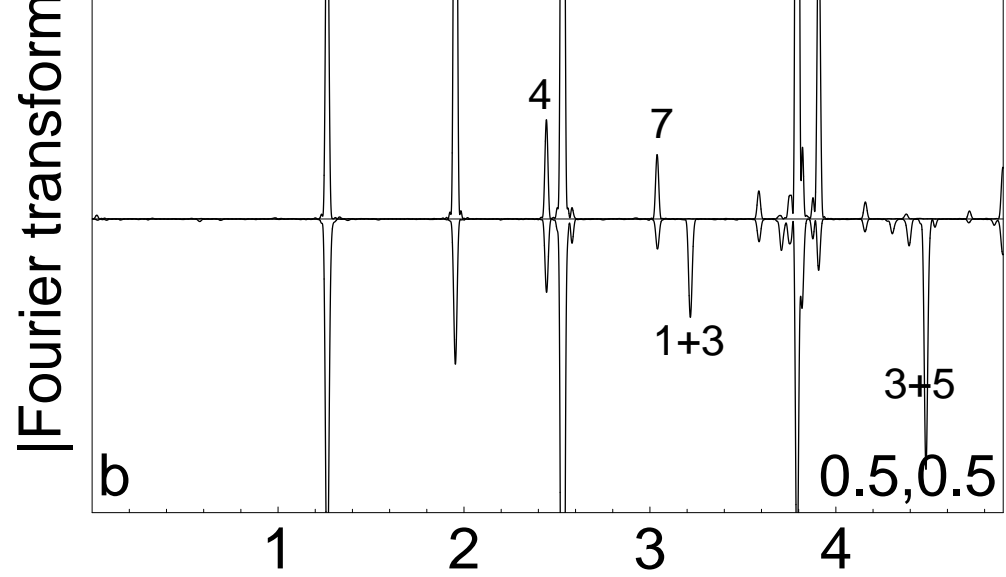
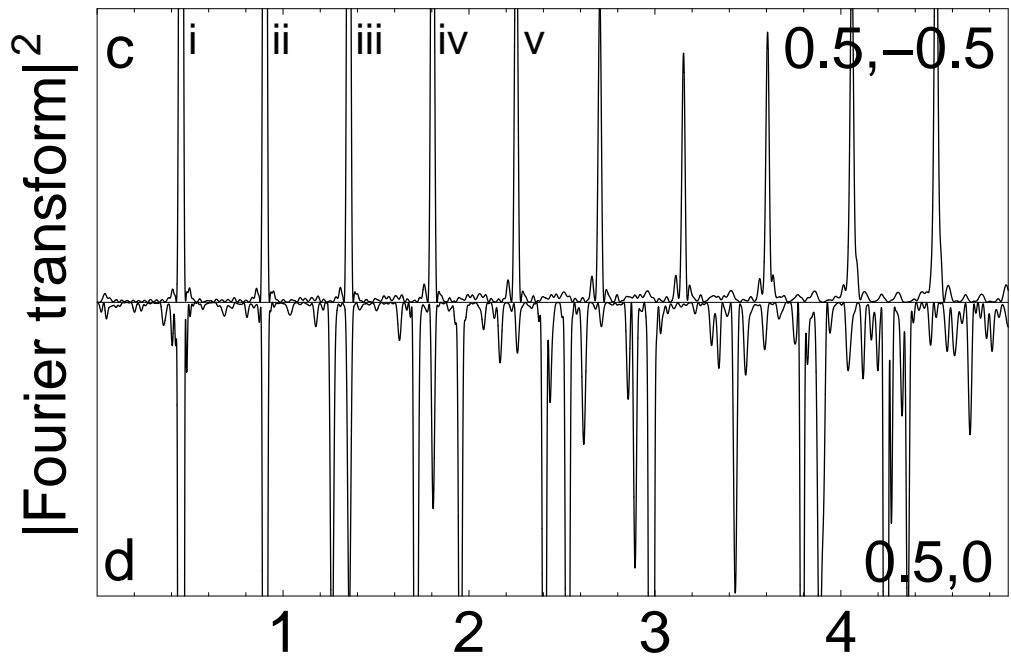
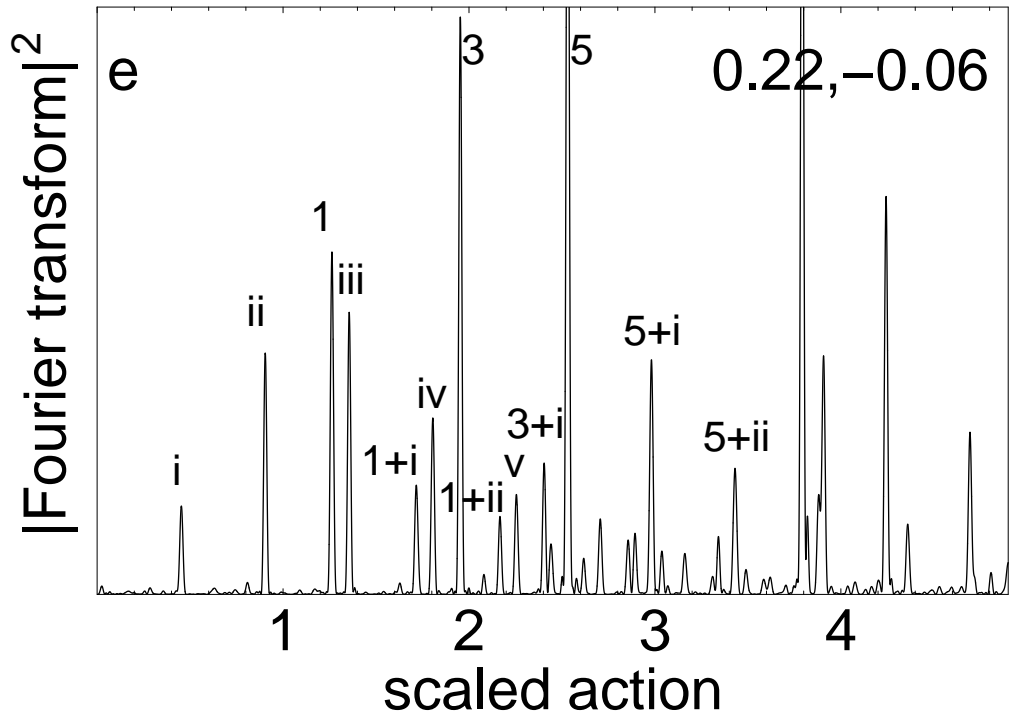
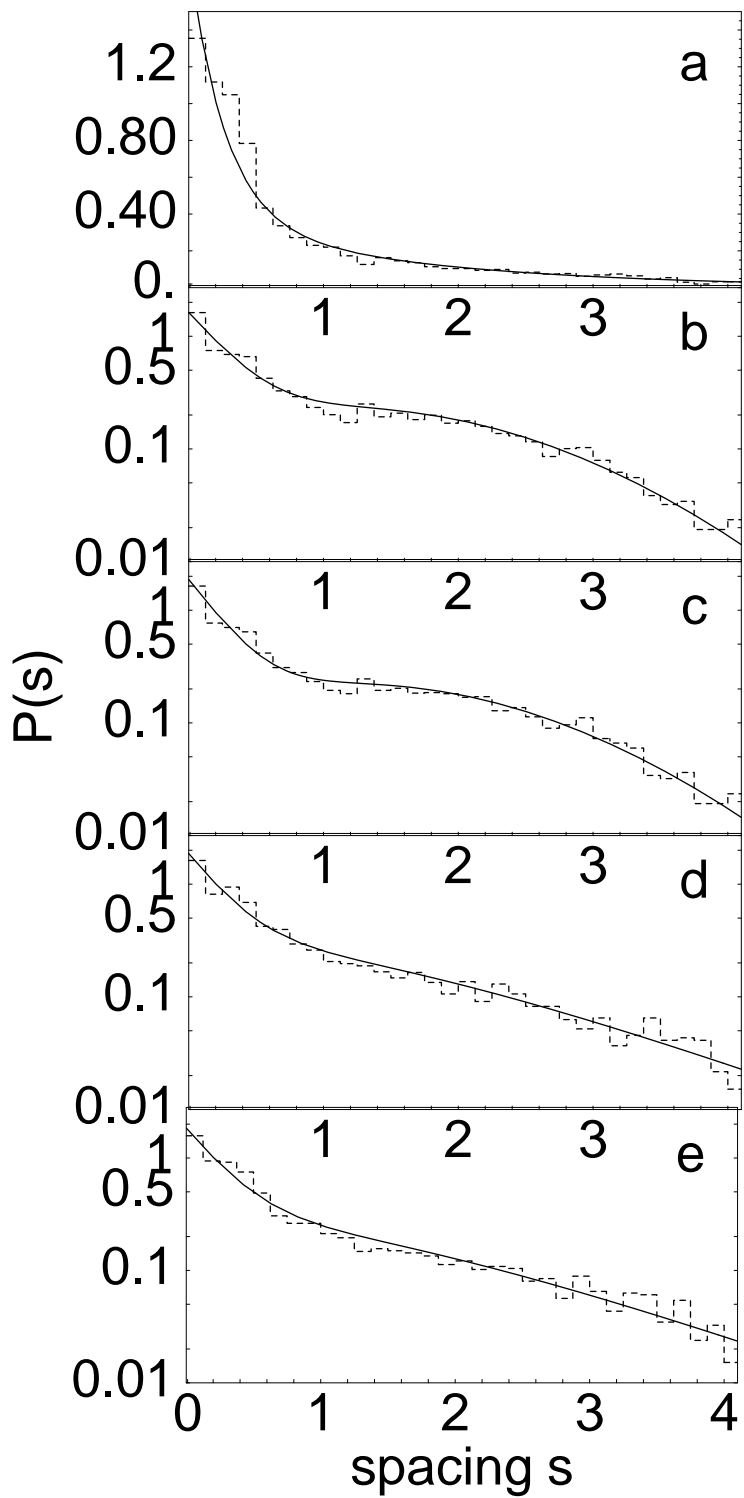
\end{document}